\documentclass[12pt]{iopart}

\usepackage{iopams}
\usepackage{graphicx}
\usepackage[breaklinks=true,colorlinks=true,linkcolor=blue,urlcolor=blue,citecolor=blue]{hyperref}

\begin{document}

\title{Tunable Thermal Conduction in Graphane Nanoribbons}
\author{Dengfeng Li$^{1,2}$, Yong Xu$^{3}$\footnote{Electronic address:yongxu@stanford.edu}, Xiaobin Chen$^{2,4}$, Bolin Li$^{1}$ and Wenhui Duan$^{2,4}$\footnote{Electronic address: dwh@phys.tsinghua.edu.cn}}

\address{$^1$Department of Mathematics and Physics, Chongqing University of Posts and Telecommunications, Chongqing, 400065, People's Republic of China}
\address{$^2$Department of Physics and State Key Laboratory of Low-Dimensional
Quantum Physics, Tsinghua University, Beijing 100084, People's Republic of China}
\address{$^3$Department of Physics, McCullough Building, Stanford University, Stanford, CA 94305-4045, USA}
\address{$^4$Collaborative Innovation Center of Quantum Matter, Tsinghua University, Beijing 100084, China}

\eads{\mailto{yongxu@stanford.edu}, \mailto{dwh@phys.tsinghua.edu.cn}}

\date{\today}

\begin{abstract}
Graphane and graphene are both two-dimensional materials but of different bonding configurations, which can result in distinct thermal conduction properties. We simulate thermal conduction in graphane nanoribbons (GANRs) using the nonequilibrium Green's function method. It is found that GANRs have lower ballistic thermal conductance and stronger thermal conductance anisotropy than the graphene counterparts. Furthermore, hydrogen vacancies of GANRs considerably suppress thermal conduction, accompanied by enhanced thermal conductance anisotropy. The tunable thermal conduction, realized by controlling the width, edge shape and hydrogen vacancy concentration of GANRs, could be useful for thermal management and thermoelectric applications.
\end{abstract}

\maketitle

\section{Introduction}

Fully hydrogenating graphene generates a chemically new two-dimensional material, named \emph{graphane}, which has $sp^3$ covalent bonds between adjacent carbon atoms, as opposed to the $sp^2$ bonds of graphene.~\cite{Sofo2007PRB153401} The hydrogenation opens the band gap, making graphane interesting for carbon-based nanoelectronics.\cite{Elias2009S610} Intensive research effort has been devoted to graphane, since it was first theoretically predicted in 2007~\cite{Sofo2007PRB153401} and experimentally realized in 2009\cite{Elias2009S610}. Previous research of graphane, mainly focusing on electronic related aspects, leads to many exciting findings\cite{Pumera2013CSR5987, Santos2013JPCC6420, Openov2011S633, Siahin2010PRB205417, Li2009JPCC15043, wang2011electronic}. However, as the size of electronic devices decreases to nanoscale, power density grows exponentially in integrated circuits,~\cite{pop2006} while, thermal conduction gets strongly suppressed due to increased numbers of boundaries and interfaces, and heat dissipation becomes serious in nanoelectronics. Therefore, more knowledge and understanding on thermal conduction behaviors are urgently demanded.

The study of thermal conduction in graphane is important not only to practical applications but also to fundamental research. It is well known that graphene belongs to one of the best thermal conductors in nature.~\cite{Balandin2011NM569} The change in bonding configuration from $sp^2$ to $sp^3$ can result in distinctly different lattice dynamics. Moreover, hydrogens of graphane prevalently desorb at finite temperatures, resulting in hydrogen vacancies that introduce phonon scattering. Importantly, concentration of hydrogen vacancy can be reversibly controlled in experiments, for instance, by varying temperature and hydrogen pressure\cite{Wang2010AN6146, Sessi2009NL4343, Ryu2008NL4597, Mikoushkin2013APL71910, Hwee2012nanoscale}. This degree of freedom of hydrogen vacancy can be used to tune thermal conduction.

In this work, we use the nonequilibrium Green's function (NEGF) method to study thermal conduction in graphane, choosing graphane nanoribbons (GANRs) as transport systems. We demonstrate that thermal conduction can be effectively tuned by controlling the width, edge shape and hydrogen vacancy concentration of GANRs. Specifically, ballistic thermal conductance $K^{\rm{ball}}$ of GANRs generally grows with the ribbon width. The scaled quantity, ballistic thermal conductance per area ($K^{\rm{ball}}/A$), displays an interesting dependence on edge shape, with zigzag nanoribbons having room-temperature $K^{\rm{ball}}/A$ up to 60\% larger than armchair nanoribbons. Similar behaviors have been found in graphene nanoribbons (GNRs).~\cite{Xu2009APL233116,tan2011nanoletter} Differently, GANRs have obviously lower $K^{\rm{ball}}/A$ and larger thermal conductance anisotropy than their graphene counterparts. Furthermore, the inclusion of hydrogen vacancies significantly enhances the thermal conductance anisotropy, and varies thermal conductance in wide ranges. The tunable thermal conduction may find applications in thermal management and thermoelectrics.

\section{Method}

Thermal conduction is contributed by electrons and lattice vibrations in graphane. The electronic part will be neglected considering the semiconducting nature of graphane. Such an approximation works well even for the gapless system graphene, in which lattice vibrations dominate thermal conduction~\cite{Balandin2011NM569}. The NEGF method\cite{Yamamoto2006PRL255503, Mingo2006PRB125402, Xu2008PRB224303, Xu2010PRB195425, Wang2008EPJB381}, as an approach fully based on quantum mechanics and widely used to deal with many-body problems, is applied to simulate thermal conduction (i.e., phonon transport herein). The complex many-body interactions, including phonon-phonon and electron-phonon interactions, would be important only at high temperatures and will not be discussed here. This allows us to focus on discussing ballistic thermal conductance and impurity scattering, for which the NEGF method gives exact results.

Here the empirical Brenner potential \cite{Brenner2002JPCM783} as implemented in ``General Utility Lattice Program"\cite{Gale1997JCSFT629} was used to relax lattice geometry and compute force constants. The optimized bond lengths of C-C and C-H of graphane are 1.54 {\AA} and 1.09 {\AA}, respectively, in good agreement with first-principles results\cite{Sofo2007PRB153401, Sahin2009APL222510, Zhou2013SSC24}. Moreover, Brenner potential is reliable for phonon calculations of carbon systems, as checked also by first-principles calculations\cite{Vandescuren2008PRB195401}. Using the obtained force constants as input, the NEGF method gives phonon transmission function $\overline{\mathcal T} (\omega)$ ($\omega$ is the phonon frequency) for a given transport system, as described in detail in our previous works~\cite{Xu2010PRB195425,zhu2012,Huang2013PRB205415,chen2013}. Then thermal conductance as a function of temperature is calculated by the Landauer formula
\begin{equation}
K(T) = \frac{k_{\rm B}^2 T}{h}  \int_{0}^{\infty} dx \frac{x^2 e^x}{(e^x -1)^2}
\overline{\mathcal T} (x),
\end{equation}
where $k_{\rm B}$ is the Boltzmann constant, $x = \hbar \omega / (k_{\rm B} T)$ with $\hbar$ the reduced Planck constant, $\overline{\mathcal T} (x) \equiv \overline{\mathcal T} (\omega)$.\cite{Xu2010PRB195425} To compare thermal conduction ability between systems of different cross sectional area ($A$), we introduced a scaled quantity, thermal conductance per area ($K/A$). For GANRs, $A = W\delta$, where $W$ is the ribbon width and $\delta$ is the effective thickness, selected to be 0.48 nm, same as the interlayer separation of graphane multilayers\cite{Rohrer2011PRB165423}.

Hydrogen vacancy induced phonon scattering was previously studied using the cascade scattering model\cite{Markussen2007PRL76803, Savic2008PRL165502, Stewart2009NL81, Wang2011APL91905} by neglecting multiple scattering induced interference effects:
\begin{equation}
\frac{1}{\overline{\mathcal T}_N}=\frac{N}{\overline{\mathcal T}_1}-\frac{N-1}{\overline{\mathcal T}_0},
\end{equation}
where $N$ is the number of hydrogen vacancies, $\overline{\mathcal T}_N$, $\overline{\mathcal T}_1$ and $\overline{\mathcal T}_0$ are the phonon transmission in the presence of $N$, 1 and 0 scatterers, respectively. The model applies to systems with low concentration of scatterers, according to previous work~\cite{Savic2008PRL165502} and our tests. All inequivalent positions of single hydrogen vacancy were considered. The corresponding phonon transmissions, showing negligible differences between each other, were averaged to give $\overline{\mathcal T}_1$. $\overline{\mathcal T}_0$ is the ballistic phonon transmission function. Once $\overline{\mathcal T}_1$ and $\overline{\mathcal T}_0$ were calculated by the NEGF method, thermal conductance as a function of $N$ was given by the cascade scattering model.

\section{RESULTS AND DISCUSSION}

We choose GANRs as transport systems, because they are building blocks of nanoelectronic devices and they are quasi-one dimensional, suitable for transport study. In experiments, GANRs can be formed by hydrogenation of GNRs, by unzipping graphane nanotubes\cite{Talyzin2011AN5132} or by directly cutting a graphane sheet.\cite{Sun2011NC559} We consider GANRs of different edge shapes, including armchair GANRs (AGANRs) and zigzag GANRs (ZGANRs). In an ideally truncated GANR, each edge carbon atom has a dangling bond. We study cases without and with hydrogen saturation at edges, with the latter case denoted by an additional label ``-H'', as depicted in Fig. 1.

\begin{figure}[tb]
\centering\includegraphics[width=0.8\linewidth]{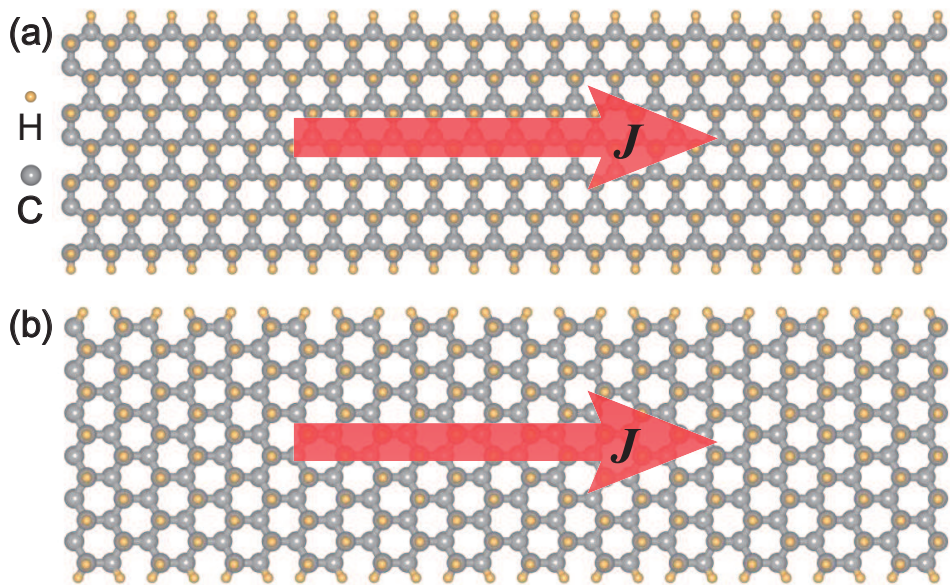}
\caption{(Color online) The geometric structures of edge H-passivated graphane nanoribbons: (a) AGANR-H and (b) ZGANR-H. The smaller/larger spheres represent hydrogen/carbon atoms. Thermal current $J$, along the longitude direction, is denoted by the red arrow.}
\end{figure}

Thermal conductivity $\kappa$ in diffusive region is given by $\kappa = \lambda K^{\rm{ball}}/A$, where $\lambda$ is the phonon mean free path for backscattering.~\cite{xu2014} The information of $K^{\rm{ball}}/A$ helps us to evaluate the ability of a material as a ballistic thermal conductor and also to estimate $\lambda$ if $\kappa$ is known from experiment. Figure 2 presents the room-temperature $K^{\rm{ball}}/A$ for GANRs of varying widths and edges. $K^{\rm{ball}}/A$ depends significantly on the width for narrow GANRs, and shows a weak width dependence for wide GANRs when $W > 1.5$ nm. Importantly, $K^{\rm{ball}}/A$ of zigzag nanoribbons is obviously larger than that of armchair nanoribbons, showing anisotropic thermal conduction. We define an anisotropy factor $\eta = (K/A)_z/(K/A)_a - 1$, where the subscripts ``$z$'' and ``$a$'' denote zigzag and armchair edges, respectively. At room temperature, $\eta$ of GANRs is as high as 60\%, gets smaller in wider ribbons, but still keeps large (over 20\%) even for ribbons of $W = 30$ nm (see inset of Fig.2). Hydrogenation at edges has minor influence on thermal conductance of zigzag nanoribbons, while noticeably increases thermal conductance of armchair nanoribbons, leading to a slight decrease of $\eta$.

\begin{figure}[tb]
\centering\includegraphics[width=0.8\linewidth]{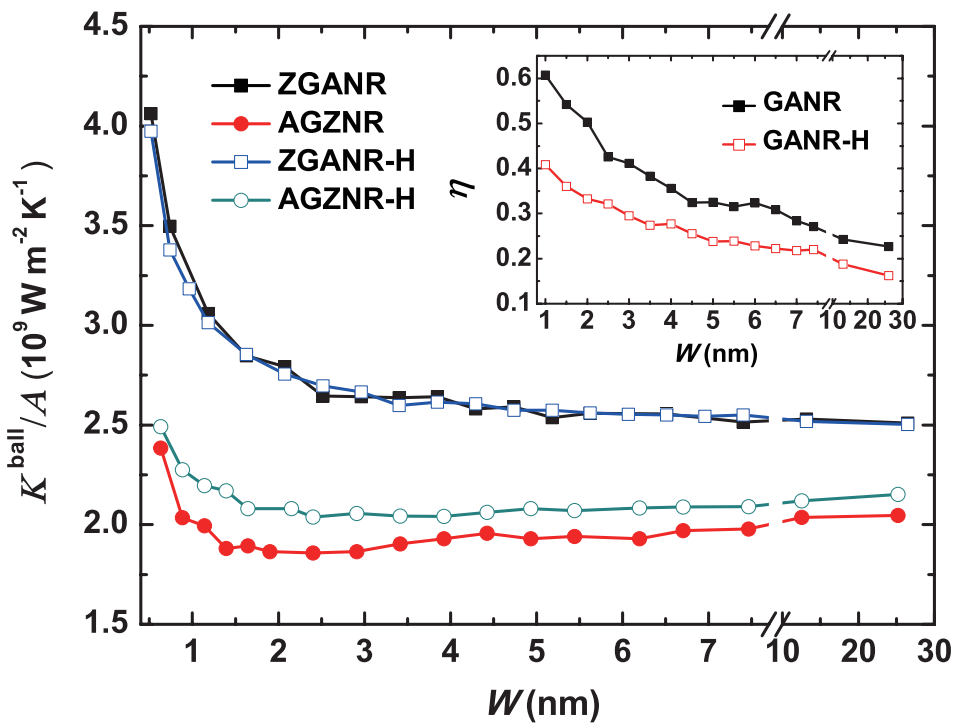}
\caption{(Color online) The scaled thermal conductance ($K^{\rm{ball}}/A$) of  GANRs with different edges as a function of width ($W$) at $T=300$ K. The inset shows the anisotropy factor $\eta$ of thermal conductance as a function of width ($W$) for GANR and GANR-H at $T =300$ K. }
\end{figure}

It is worthwhile to compare ballistic thermal conductance between graphane and graphene systems. At room temperature and for $W \sim 10$ nm, $K^{\rm{ball}}/A$ of zigzag GNR is $4.2\times 10^9$ Wm$^{-2}$K$^{-1}$,~\cite{Xu2009APL233116} close to that of graphene sheet~\cite{xu2014}. In contrast, the corresponding value of ZGANRs is $2.5\times 10^9$ Wm$^{-2}$K$^{-1}$, about 40\% lower than the graphene counterpart. When excluding the influence of a larger effective thickness, we still find a decrease of 15\% in $K^{\rm{ball}}/W$ from graphene to graphane cases. The decrease of ballistic thermal conductance is presumably because phonon bands of graphene is more dispersive than those of graphane. This is consistent with previous studies~\cite{Pei2011C4752,Kim2012AN9050, Liu2012APL71910}, which find reduced phonon group velocity when changing the bonding configuration from $sp^2$ to $sp^3$.

Similar anisotropic thermal conductance is also found in GNRs~\cite{Xu2009APL233116, tan2011nanoletter, aksamija2011apl, Huang2013PRB205415}, where the room-temperature $\eta$ is up to 30\%, smaller than that of GANRs. Considering that thermal conduction is isotropic in a two-dimensional graphane or graphene sheet, the appearance of anisotropy in nanoribbons can be attributed to boundary effects. As shown in the inset of Fig. 2, $\eta$ keeps noticeably large even for nanoribbons of tens of nanometers wide, indicating that the boundary effects are very long range.

Intuitively phonon structure of the same material would vary if applying different boundary conditions, which contributes to the anisotropy in $K^{\rm{ball}}/A$. Previous work~\cite{Huang2013PRB205415} argued that in hexagonal lattice structures armchair edges are prone to give more localized phonon modes than zigzag edges, resulting in lower phonon transmission. The argument applies to present systems, as supported by comparing phonon transmission between ZGANR and AGANR with similar width of 3.4 nm. As shown in Fig. 3(a), phonon transmission of ZGANR is obviously higher than that of AGANR, especially within the frequency region of 200-700 cm$^{-1}$. Thermal conductance, as a weighted integration of phonon transmission function, is thus larger in ZGANR than in AGANR. Meanwhile, we notice that phonon transmissions of the two GANRs are  essentially the same in the low-frequency region (below 200 cm$^{-1}$). The low-temperature thermal conductance, mainly contributed by low-frequency phonons, should be isotropic. As presented in the inset of Fig. 3(b), $\eta$ evolves from 0 at low temperatures, gradually increases and finally saturates when increasing temperature to above 200 K.

\begin{figure}[tb]
\centering\includegraphics[width=0.8\linewidth]{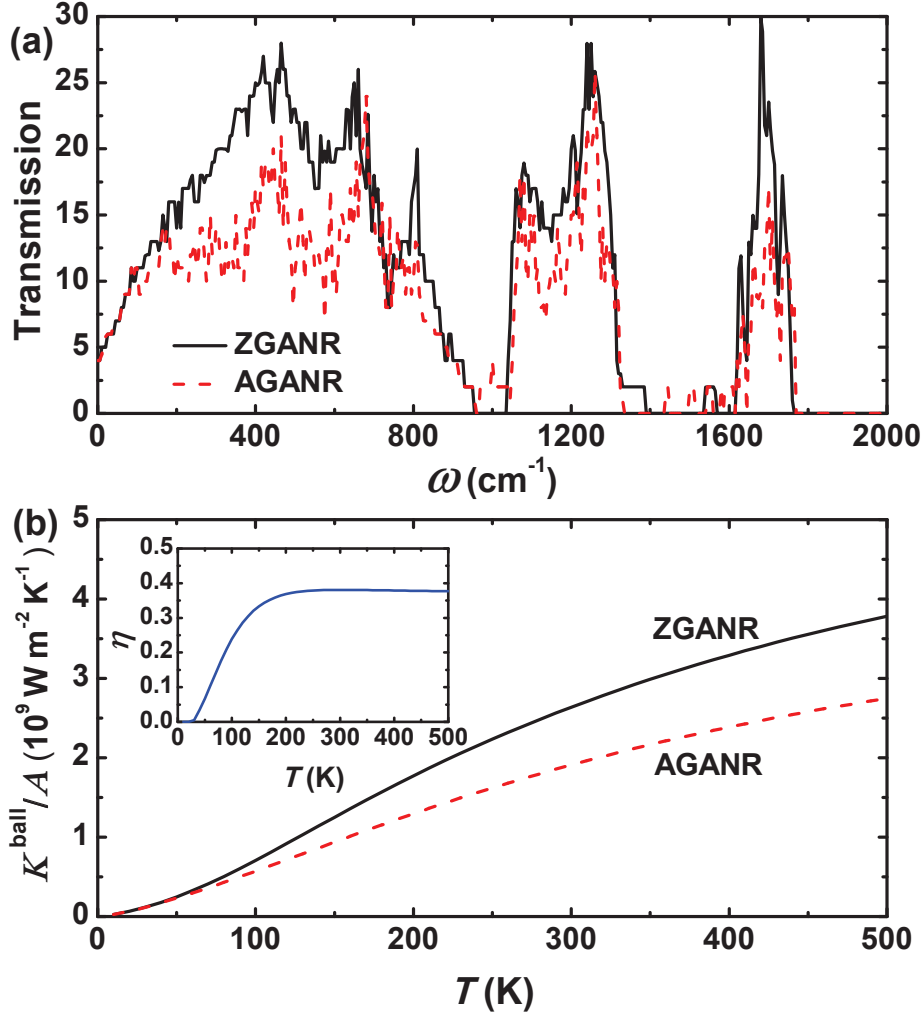}
\caption{(Color online) (a) Transmission function versus phonon frequency ($\omega$) and (b) the scaled thermal conductance ($\kappa^{\rm{ball}}/A$) versus temperature ($T$) for ZGANR and AGANR of width $W=3.4$ nm. The inset shows the anisotropy factor $\eta$ as a function of temperature $T$.}
\end{figure}

In graphane, hydrogen vacancies commonly exist and their concentration can be effectively controlled by experimental conditions like temperature and hydrogen pressure.\cite{Wang2010AN6146, Sessi2009NL4343, Ryu2008NL4597, Mikoushkin2013APL71910,Hwee2012nanoscale} It would be interesting to see the effect of hydrogen vacancy on thermal conduction. Physically each hydrogen vacancy acts as an impurity scatterer for phonon transport. Thermal conductance gets lower with increasing number of hydrogen vacancies ($N$), as shown in Fig. 4(a). This can change thermal conductance orders of magnitude when $N$ is large. The $N$-dependent thermal conductance, if providing a concentration of hydrogen vacancy, naturally gives the length-dependent behavior, which enables extracting phonon mean free path of impurity scattering. Further details will be presented elsewhere. Though hydrogen vacancies decrease thermal conductances of both ZGANR and AGANR, the influence is quantitatively different. We find that thermal conduction reduction induced by hydrogen vacancy is smaller in ZGANR than in AGANR, leading to enhanced thermal conductance anisotropy. As shown in Fig. 4(b), $\eta$ enhances from 40\% to 65\% when increasing $N$.

\begin{figure}[tb]
\centering\includegraphics[width=0.8\linewidth]{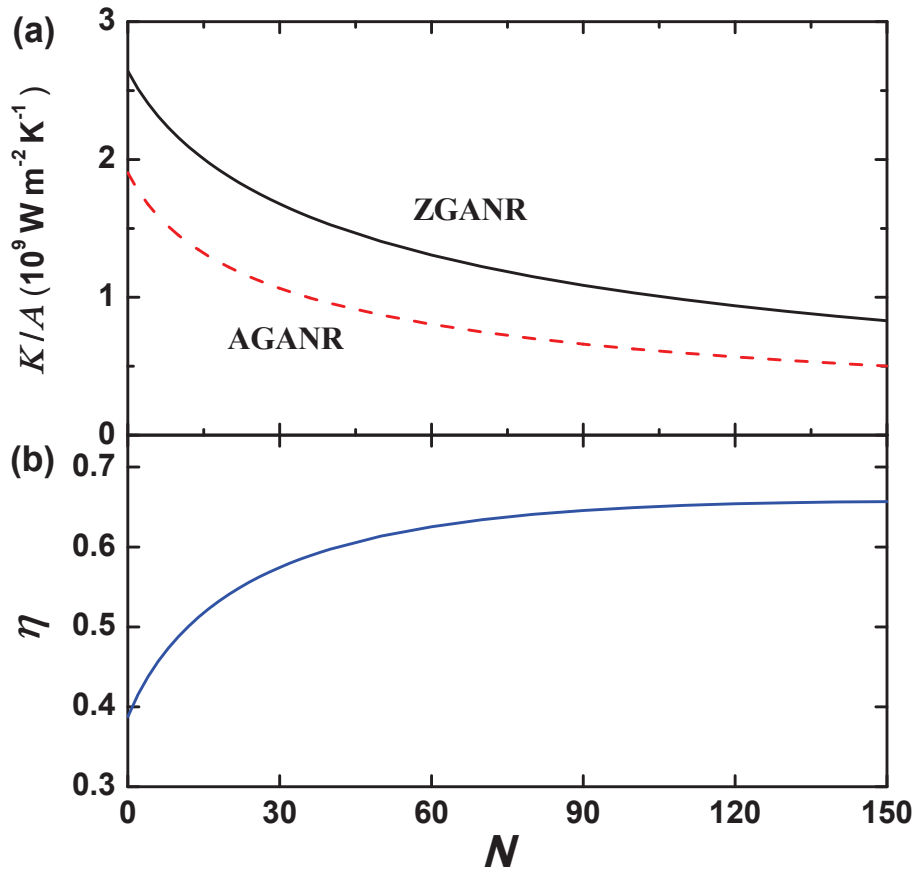}
\caption{(Color online) (a) The scaled thermal conductance ($\kappa/A$) versus the number of H-vacancy $N$ for ZGANR and AGANR of width $W=3.4$ nm at $T=300$ K. (b) The anisotropy factor $\eta$ versus the number of H-vacancy $N$ at $T=300$ K. }
\end{figure}

To demonstrate hydrogen vacancy induced scattering in detail, we present phonon-transmission ratios between systems with and without hydrogen vacancies, as shown in Fig. 5. A single hydrogen vacancy does not affect low-frequency acoustic phonons and mainly scatters high-frequency optical phonons, in qualitative agreement with the picture of Rayleigh scattering. As the number of vacancies increases, phonon transmission gets suppressed (except for phonons of nearly zero frequency). The trend is independent of edge shapes of GANRs. However, hydrogen vacancy induced suppression of phonon transmission is obviously weaker in ZGANR than in AGANR. This provides an example that the variance in boundary condition affects the strength of impurity scattering. The difference in scattering strength is translated to the discrepancy in thermal conductance reduction, which explains the enhancement of thermal conduction anisotropy induced by hydrogen vacancy.

\begin{figure}[tb]
\centering\includegraphics[width=0.9\linewidth]{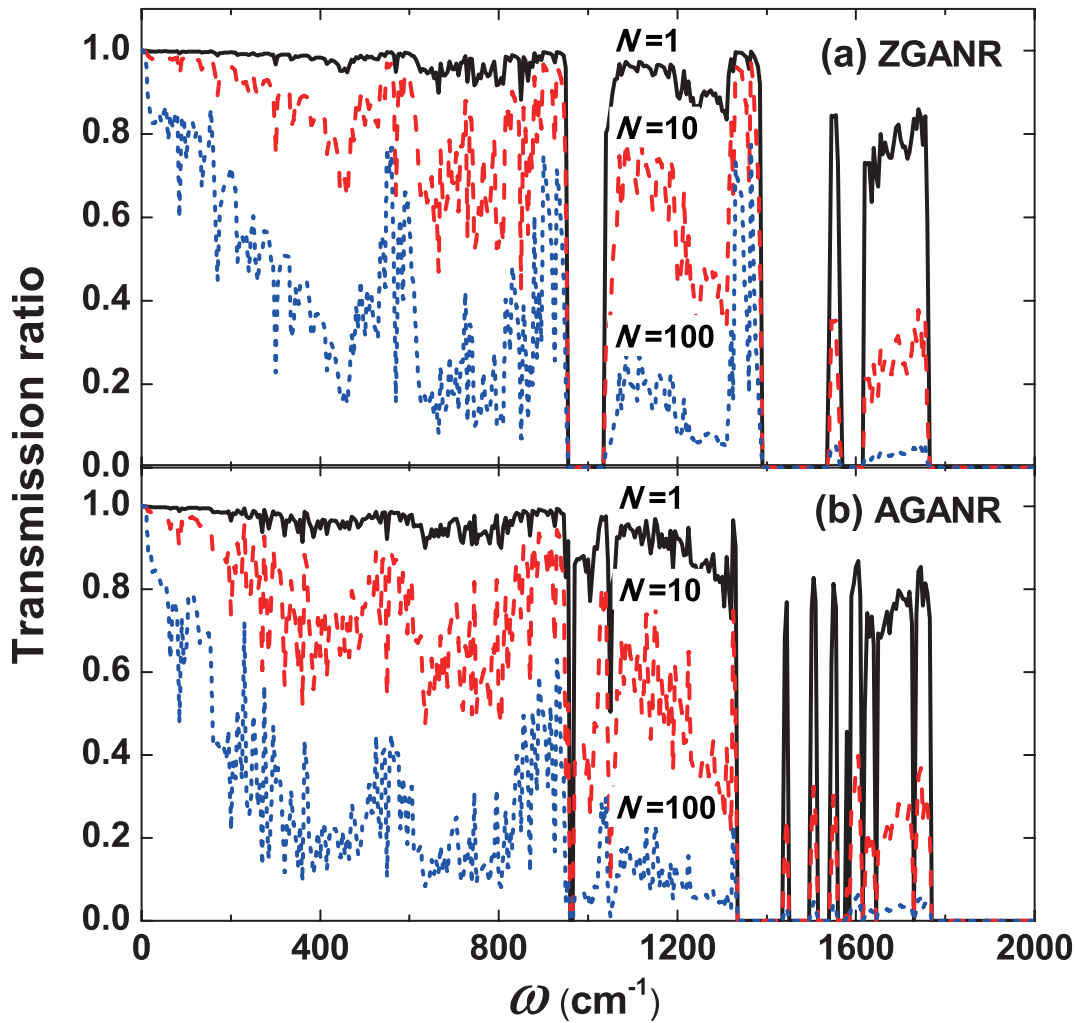}
\caption{(Color online) The phonon transmission ratio to the ballistic limit for (a) ZGANR and (b) AGANR of $W=3.4$ nm with $1$, $10$ and $100$ H-vacancies.}
\end{figure}

\section{CONCLUSIONS}

In summary, we show dependencies of thermal conductance on the width, edge shape and hydrogen vacancy concentration of GANRs. These degrees of freedom allow tuning thermal conduction in graphane systems. Compared to the graphene counterparts, GANRs are expected to have lower thermal conductivity, because their ballistic thermal conductance is lower due to the different bonding configuration, and their phonon mean free path in GANRs is shorter caused by hydrogen vacancies. It is known that graphene is not suitable for thermoelectrics because of its gapless band structure and high thermal conductivity.~\cite{xu2014} Both disadvantages, however, are overcome in graphane, offering great promise for thermoelectrics.

\ack We acknowledge the support of the Ministry of Science and Technology of China (Grants No. 2011CB921901 and 2011CB606405), the National Natural Science Foundation of China (Grants No. 11074139 and 11334006), and the Key Program of the Natural Science Foundation of Chongqing of China (Grant No. cstc2012jjB50010).

\section*{References}
\providecommand{\newblock}{}

\end{document}